# Number of Sunspot Groups from the Galileo-Scheiner controversy revisited

V.M.S. Carrasco[1,2], M.C. Gallego[1,2], J.M. Vaquero[2,3]

[1] Departamento de Física, Universidad de Extremadura, 06071 Badajoz, Spain [e-mail: vmscarrasco@unex.es]

[2] Instituto Universitario de Investigación del Agua, Cambio Climático y Sostenibilidad (IACYS), Universidad de Extremadura, 06006 Badajoz, Spain

[3] Departamento de Física, Universidad de Extremadura, 06800 Mérida, Spain

**Abstract:** We revise the sunspot observations made by Galileo Galilei and Christoph Scheiner in the context of their controversy on the nature of sunspots. Their sunspot records not included in the current sunspot group database, used as a basis to calculate the sunspot group number, are analyzed. Within the documentary sources consulted in this work, we can highlight the sunspot observations by Scheiner included in the letters sent under the pseudonym *Apelles* to Marcus Welser and the first sunspot observations made by Galileo, which can be consulted in *Le opere di Galileo Galilei*. These sunspot observations would extend the temporal coverage for these two observers and filling some gaps in the current group database in the earliest period where the data available is sparse. Moreover, we have detected changes in the quality of the sunspot drawings made by Galileo and Scheiner in their observation series affecting to the number of groups recorded by the two observers. We also compare these records with sunspot observations made by other astronomers of that time. According to this comparison and regarding the same observation days, Scheiner was generally the astronomer who reported more sunspot groups while Harriot, Cigoli, and Galileo recorded a similar number of groups. We conclude these differences are mainly because of the observational method used by the observers.

**1. Introduction**

The sunspot number index is a measure of solar activity calculated from the number of sunspots observed on the solar photosphere (Clette & Lefèvre 2016; Hoyt & Schatten 1998). This is the most used index to study long-term solar activity and it is employed in studies of different scientific fields such as solar physics or Earth sciences (Kopp et al. 2016; Usoskin 2017). Thus, its reliability is fundamental. Recently, several corrections were applied to the sunspot number to fix scaling problems in the 19th and 20th centuries (Clette et al. 2014) and an ongoing global effort is carried out in order to provide a recalibration of these indices (Muñoz-Jaramillo & Vaquero 2019).

Sunspots have been observed before the telescopic era (Yau & Stephenson 1988; Vaquero, Gallego & García 2002). However, the systematic record of sunspot observations did not start until the use of the telescope as an astronomical instrument (Vaquero & Vázquez 2009; Arlt & Vaquero 2020). The first sunspot record available of the telescopic era was made by Thomas Harriot on 18 December 1610 (all dates in this work are referred to the Gregorian calendar). Since this record, several astronomers made sunspot observations in those first years of telescopic observations. According to the current sunspot group database (Vaquero et al. 2016, hereafter V16), the observers with the highest number of sunspot records in the two first decades of telescopic era (1610 – 1629) were Christoph Scheiner (766 observation days), Thomas Harriot (210), Charles Malapert (187), Daniel Mögling (126), and Joachim Jungius (104). It is important to note that the observational coverage in this early period is low. For the period 1610 – 1629, the observational coverage is around 20 % according to V16 and if we only consider the first decade, from December 1610 to 1619, it is even lower (15 %). For this reason, the incorporation of new information on sunspot observations made in those first years of the telescopic era is fundamental to understand how solar activity evolved in that period. We highlight that, recently, Neuhäuser & Neuhäuser (2016), Carrasco (2019), and Carrasco et al. (2019a) found some problems in the sunspot counting from the records of the first years of telescopic observations. Furthermore, Carrasco et al. (2019b) found by revisiting Malapert records, on the one hand, the two first solar cycles of the telescopic era had the shape of the standard 11-year solar cycle and, on the other hand, the solar activity level calculated from Malapert (1633) was one-third greater than that obtained from V16 considering the same documentary source. Recently, Vokhmyanin, Arlt & Zolotova (2020) calculated sunspot areas and positions from Harriot's drawings. In this context, it is also interesting to revisit the sunspot observations made by astronomers of that time like, for example, Galileo and Scheiner. We also note that these first observations precede the Maunder Minimum (Eddy 1976; Usoskin et al. 2015), the period between 1645 and 1715 characterized by a long and prolonged low solar activity, which makes difficult to connect the first data with modern observation series (Muñoz-Jaramillo & Vaquero 2019). Thus, this gives a significant importance to any new sunspot data series from that early era.

There are 51 observation days by Galileo Galilei in V16. Although he did not record many sunspot observations, Galileo was an important sunspot observer. He is the fourth most active sunspot observer if we consider records made during the first decade of telescopic era. Galileo indicated that he observed sunspots in 1610 (Galilei 1895a) and therefore he could be the first astronomer who observed sunspots by telescope but he did not record the exact date. Furthermore, he made very detailed sunspot drawings from May to August 1612 and had a great debate with Scheiner about the nature of sunspots (Sakurai 1980; Galilei & Scheiner 2010). While Scheiner defended the Aristotelian ideas about a perfect Sun, Galileo demonstrated that sunspots are on the solar "surface" (today photosphere). The sunspot observations involved in this famous controversy of the history of the astronomy have allowed us to know the level of solar activity in those first years of the telescopic era. Although these sunspot reports are widely known in the context of this discussion, some of them are not included in V16. Christoph Scheiner was also one of the most important sunspot observers of that time because of his number of sunspot records and the outstanding quality of his drawings published in *Rosa Ursina* (Scheiner, 1630). The sunspot observations included in this documentary source were made in 1620s, except one of the sunspot drawings made by Malapert (in 1618). We highlight that Casas et al. (2006) estimated the solar differential rotation from the Galileo's drawings and, more recently, Arlt et al. (2016) and Vokhmyanin & Zolotova (2018) calculated the sunspot positions from the observations published by Scheiner from 1611 to 1631 and Galileo from May to August 1612, respectively. Figure 1 represents the daily number of sunspot groups recorded from 1610 to 1618 according to V16, including the sunspot observations made by Scheiner (red bars) and Galileo (blue bars). We have selected, among the different countings from Galileo's observations in V16, the observer with the station number 3 because it has the greatest observational coverage for this observer.

In this work, we analyze the sunspot observations made by Scheiner from December 1611 to August 1613 and Galileo from February to May 1612. Surprisingly, despite of the fame of these astronomers, these observations are not included in the last revised collection of sunspot group numbers by V16 nor in the previous version (Hoyt & Schatten 1998). We describe the sunspot observations in Section 2. Section 3 includes the sunspot counting carried out from these observations and a comparison to the sunspot observations carried out by other astronomers of that time. Finally, the main conclusions of this work are shown in Section 4.

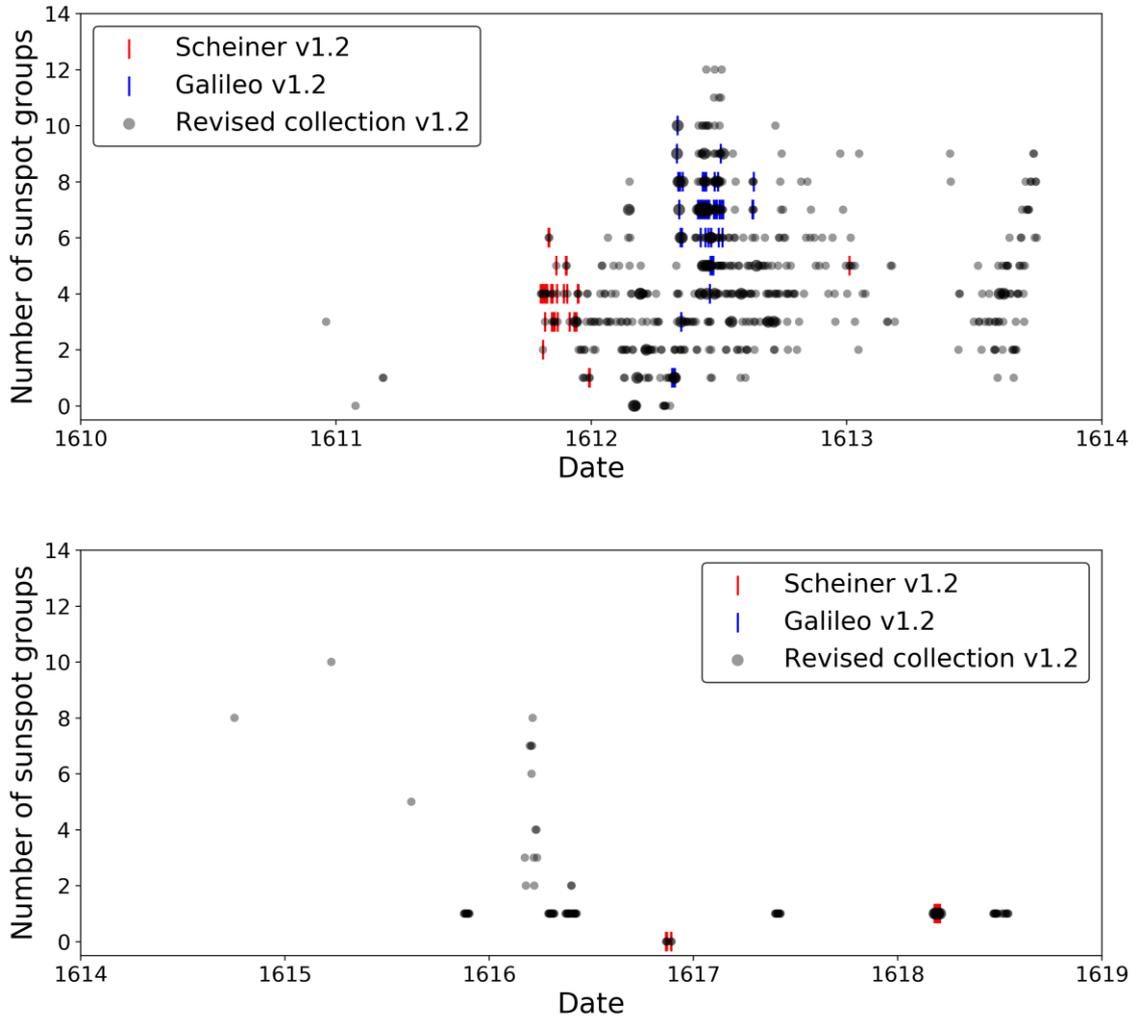

Figure 1. Daily number of sunspot groups recorded by Scheiner (red color) and Galileo (blue color) included in V16, and all the sunspot records available in V16 (grey color) during the period 1610 – 1613 (top panel) and 1614 – 1618 (bottom panel). The sizes of dots are function of the number of identical values recorded by all the observers in the same date such that greater sizes depict greater number of observations per day.

**2. Documentary sources and observations**

Christoph Scheiner recorded sunspot observations from 1611 to 1640 and he is clearly the observer with the highest number of sunspot records before the Maunder Minimum (Arlt et al. 2016; Engvold & Zirker 2016; Vaquero et al. 2016). Scheiner (1630) published a great number of drawings including his own sunspots records and sunspot observations made by other astronomers as, for example, Charles Malapert (Carrasco et al. 2019b). In particular, the sunspot drawings made by Scheiner in 1620s have extraordinary quality due to its high level of detail. Thus, '*Rosa Ursina*' could be considered the main documentary source of observations and discussion about sunspots regarding the first years of the telescopic era.

The first sunspot drawing set recorded by Scheiner includes observations made from 21 October 1611 to 14 December 1611. Another very similar series of observations was published by Scheiner (1612) and includes his sunspot records made from 10 December 1611 to 11 January 1612 and for March–April 1612. They are part of the letters Scheiner sent to Marcus Welser signed under the pseudonym *Apelles*. These drawings can be consulted, for example, on the website of the Lincean Academy Archive (https://bibdig.museogalileo.it). Furthermore, four sunspot drawings made in 1612 (23 October, 27 and 29 November, and 26 December) and one more in 1613 (6 January 1613) were published by Scheiner (1615). Other sunspot drawing made by Scheiner on 1 August 1613 was sent by Welser to Galileo in a letter written in October 1613 (Galilei 1895b). Figure 2 includes all these observations, except ten common observation days with Galileo and other common observation day with Colonna on 1 August 1613. These last drawings are represented in other figures (Figure 5, 6, and 7) to reduce the extension of Figure 2. Note that the original disc diameter of the sunspot observations made by Scheiner between 1611 December 14 to 1612 April 7 is about 22.7 mm (Arlt et al. 2016). Thus, 0.1 mm on the discs corresponds to 0.5º in heliographic coordinates in the disc centre. These sunspot records made by Scheiner were not included in V16. We must point out here that the level of detail included in the sunspot drawings made by Scheiner from 1611 to 1613 is significantly lower than in the sunspot drawings published by Scheiner (1630) in *Rosa Ursina*. For example, sunspots are represented by simple dots without penumbra in these first drawings while complex formation of sunspots and facular regions can be found in *Rosa Ursina*. We note that a transit of Venus in front of the Sun was expected on 11 December 1611 according to some predictions computed using the Ptolemaic system. Scheiner (1615) represented his estimation for Venus' size in the four sunspot drawings made at the end of 1612 (dot "A") in order to compare its shape and size with those of the sunspots recorded those days. Scheiner estimated the apparent diameter for Venus in 3 minutes of arc (Galilei & Scheiner 2010, p. 66). However, Arlt & Vaquero (2020) indicated the expected size of Venus provided by Scheiner in those drawings is actually a bit smaller than half that size in its lower conjunction and, moreover, no Venus' transit occurred but an upper conjunction.

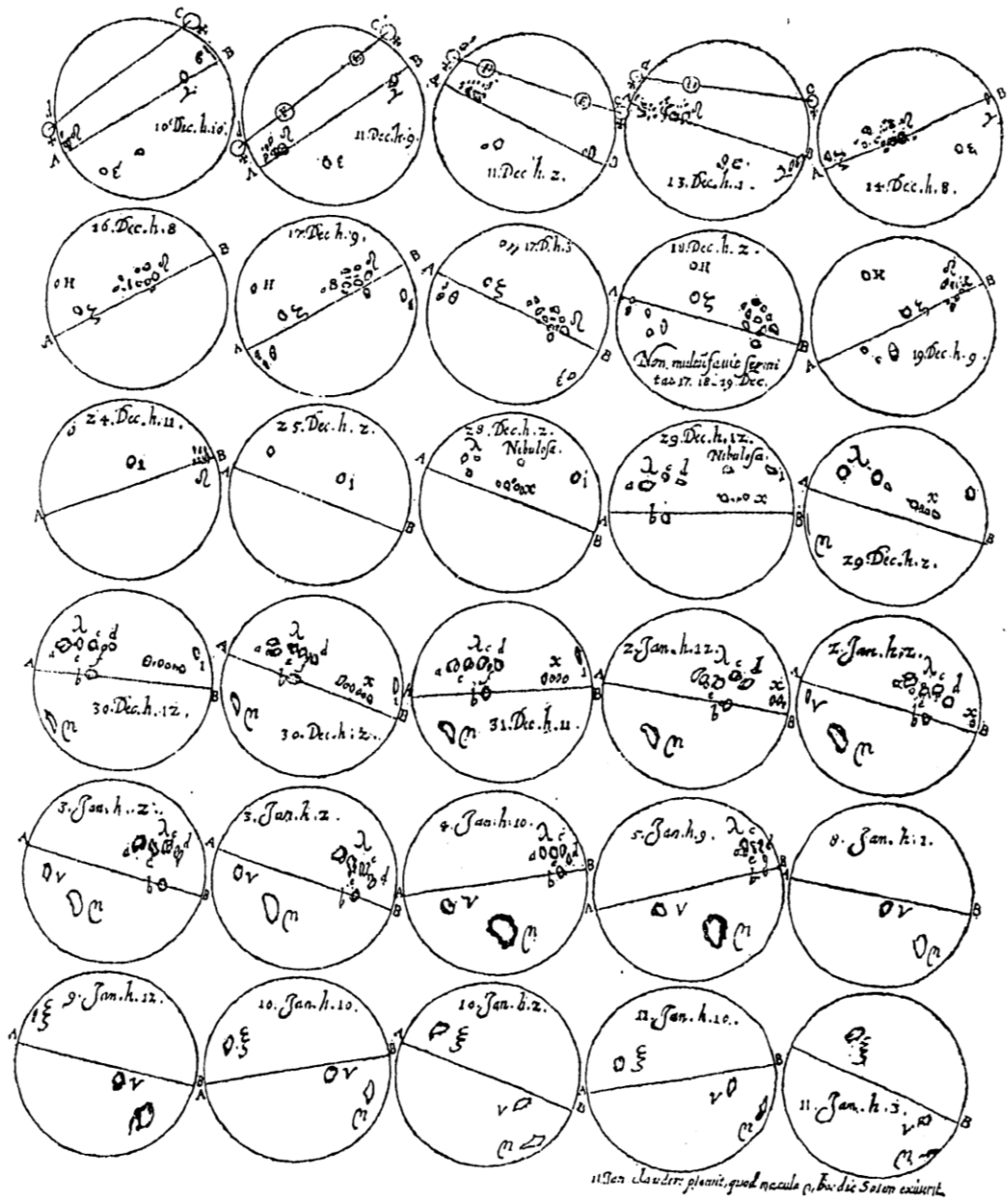

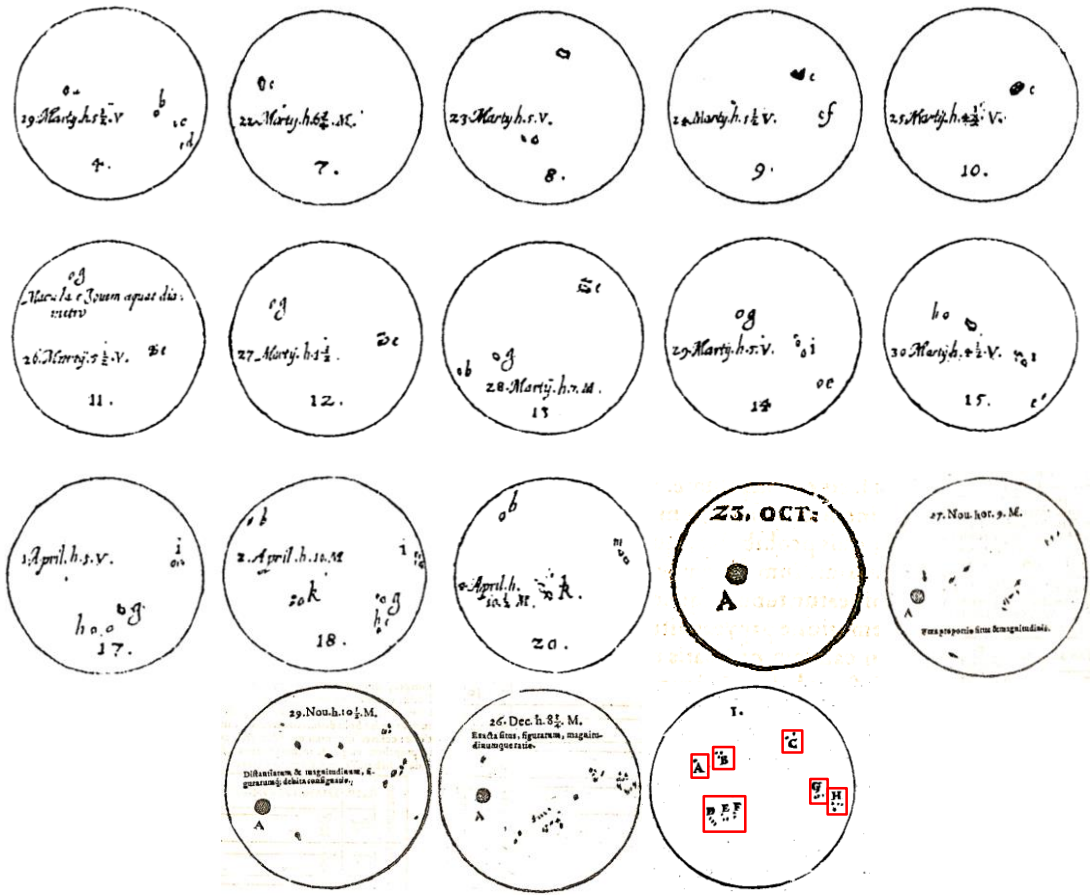

Figure 2. Sunspot observations made by Scheiner from 10 December 1611 to 6 January 1613. Note that: (i) dot "A" in the drawings corresponding to 23 October, 27 and 29 November, and 26 December 1612 represents the planet Venus, and (ii) dot "I" on the last drawing (6 January 1613) does not represent a sunspot and red squares depict different sunspot groups according to this work (6 in this case) in order to show an example of how groups have been counted [Source: Galilei 1895a; Scheiner 1615].

Galileo Galilei mentioned that he observed sunspots from 1610 although there are no exact dated records until 1612 (Galilei 1895a). Instead, he published his sunspot drawings made from February to August 1612. The drawings corresponding to June–August were published in *Istoria e dimostrazioni intorno alle macchie solari* (Galilei 1613) and those for 3–11 May 1612 are included in the letters sent by Galileo to Cardinal Maffeo Barberini (later Pope Urban VIII) whose original manuscripts are preserved at the Vatican Library (Vokhmyanin & Zolotova, 2018). These observations represent the most detailed sunspot drawings made in the first ten years of the telescopic era and the group counting from them is included in V16. However, most of the sunspot observations made by Galileo before 3 May 1612 were not taken into account in V16. These sunspot drawings can be consulted in the *Frammenti attenenti alle lettere sulle macchie solari* included in *Le opere di Galileo Galilei* (Galilei 1895a). Furthermore, textual descriptions are included in these drawings and even information about spotless days (2 and 4 March 1612) can be extracted. Galileo also reported in a comment that he observed sunspots on 1 March 1612, but he did not publish the sunspot drawing corresponding to this date and did not specify the number of groups or single sunspots. These observations made by Galileo but missing in V16 are shown in Figure 3.

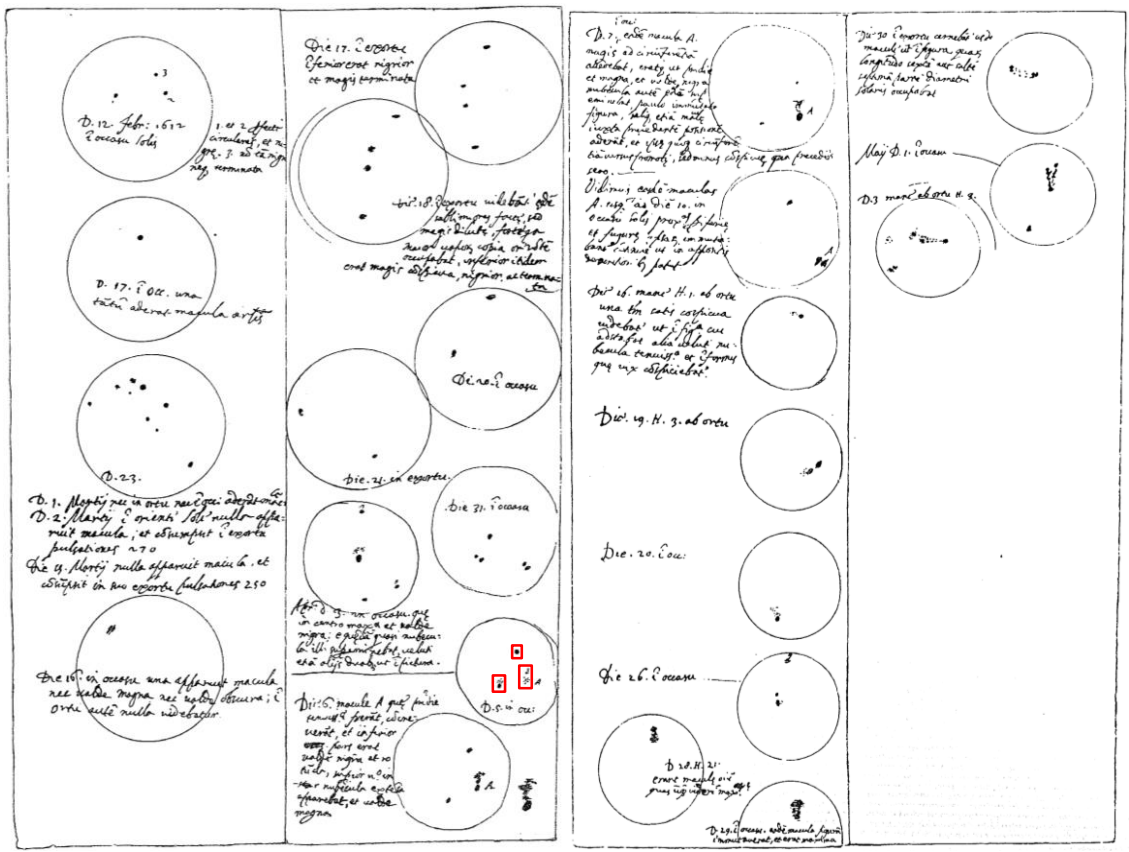

Figure 3. Sunspot observations made by Galileo from 12 February 1612 to 3 May 1612. Red symbols depicted on the drawing corresponding to 5 April 1612 are an example to show how groups (3 in this case) have been counted in this work [Source: Galilei 1895a].

## 3. Analysis and discussion of the sunspot records

3.1. The sunspot counting

We have analyzed the documentary sources that include sunspot observations which were not incorporated to V16. Thus, the total number of records studied in this work has been 77. Regarding these records, 51 were performed by Scheiner and 26 by Galileo. All the observations made by Galileo were carried out in 1612. We have analyzed 14 records made by Scheiner in December 1611 in addition to his 61 observations in 1612 and 2 in 1613. Figure 4 (top panel) depicts the sunspot counting from October 1611 to October 1613 according to the sunspot observations made by Galileo (cyan color) and Scheiner (orange color) analyzed in this work, and all the daily sunspot group numbers included in V16 (black color) highlighting the sunspot records made by Galileo (blue color) and Scheiner (red color) included in V16. Through the observations analyzed in this work, we can fill some gaps in V16 in days where there is no information, namely, 3 days in December 1611 (16, 19, and 28 December 1611) and 15 days in 1612 (2–3, 8–10 January, 25 and 29–30 March, 2, 5 and 19–20 April, 23 October, 29 November, and 26 December 1612). In this way, the number of days with sunspot records would increase from 48 to 51 in 1611 and from 251 to 266 in 1612 in V16. Figure 4 (bottom panel) represents a comparison for the period October 1610 – January 1613 between the monthly number of observation days according to V16 (purple) and the same but including the sunspot observations analyzed in this work (green). We have represented data until January 1613 in Figure 4 (bottom panel) because from January to August 1613 (last observation analyzed in this work) no gaps would be filled in V16 from this work. In particular, for each observer, the total number of sunspot observations recorded by Galileo would increase from 51 to 72 and the amount of sunspot records made by Scheiner from December 1611 to August 1613 would increase from 33 to 78. In this work, we have counted sunspot groups by searching similar distributions to modern group classifications (McIntosh 1990). Examples of our group counts can be seen in Figure 2 (6 January 1613) and Figure 3 (5 April 1612) marked by red squares. We note that the sunspot group counting in the historical observations is sometimes not trivial and the interpretation of what is a sunspot group can slightly vary from author to author. Moreover, the number of sunspot groups included in V16 on 13, 29 and 30 December 1611 corresponding to Scheiner's observations and on 26 April 1612 regarding Galileo's records must be corrected according to this new analysis. This new counting is publicly available on the website of the Historical Archive of Sunspot Observations (HASO, haso.unex.es).

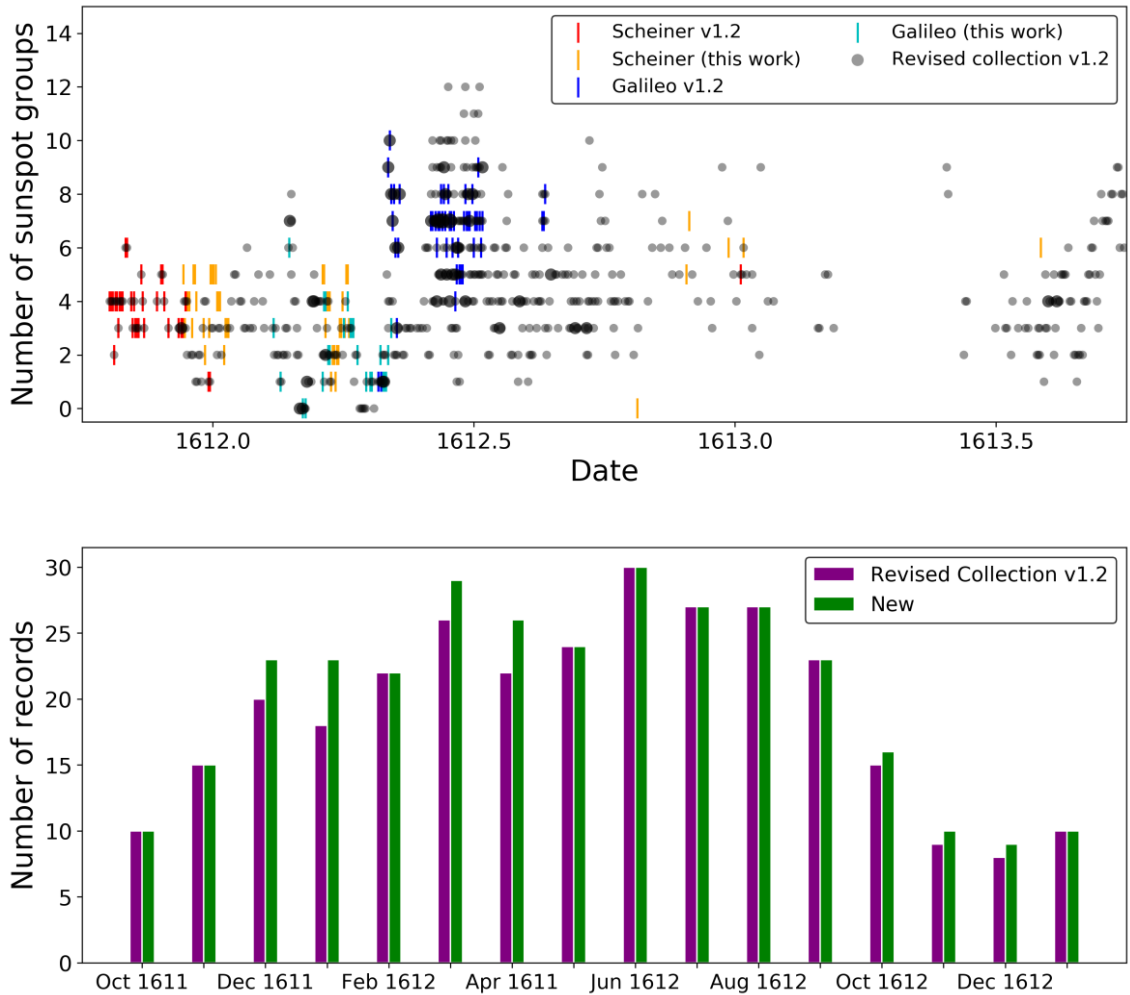

Figure 4. (Top panel) Daily number of groups recorded by Galileo (cyan color) and Scheiner (orange color) analyzed in this work. The daily group counting included in V16 for the period October 1610 – October 1613 according to Galileo and Scheiner, in addition to the entire sunspot observation set are represented in blue, red, and black color respectively. The sizes of dots are function of the number of identical values recorded by all the observers in the same date such that greater sizes depict greater number of observations per day. (Bottom panel) Monthly number of observation days according to V16 (purple) and that including the sunspot observations analyzed in this work in V16 (green) for the period October 1610 – January 1613.

3.2. Comparing sunspot observations

By comparing all the sunspot observations made by Galileo analyzed in this work with those already included in V16, we realized of an important fact. There is a significant difference between two sunspot drawings made by Galileo on 3 May 1612. The number of sunspot groups recorded by Galileo for this date according to the sunspot drawing analyzed in this work is three. It is significantly lower than the nine groups recorded by Galileo according to V16 for the same day. This last value agrees with the observation corresponding to 3 May included in a more detailed sunspot drawing set made by Galileo from 3 to 11 May 1612 (Vokhmyanin & Zolotova, 2018). One difference between both observations is that the sunspot drawing with three groups was made after the sunrise while the observations where we can see nine groups were made in the afternoon (see page 82 in Galilei & Scheiner 2010). Thus, in the drawing made after sunrise, we can see that only the two largest groups and other big spot were recorded. However, the most remarkable difference between both records was because of the change in the observation method by Galileo. The first sunspot observations made by Galileo were carried out by telescope, for example, through the clouds. Benedetto Castelli, student of Galileo, discovered that one could project the solar image to observe the Sun in anytime during the day and not only behind clouds (Galilei & Scheiner 2010). Castelli completed this method projecting the Sun on a sheet that included a predetermined size of the solar disc, adjusting the solar disc to the circle of the sheet. Thus, Galileo made the first drawing on 3 May 1612, just after the sunrise, observing sunspots through the telescope and the drawing at the end of that day using the new methodology by projecting the solar image in a circle of 12.5 cm. The level of detail between these two drawings show the transition from old to new methodology and gives an idea on how affected to the number of sunspot groups recorded in the observation series by Galileo.

Table 1. Comparison between the daily number of sunspot groups obtained from the sunspot drawings made by Galileo and Scheiner analysed in this work for the same observation days.

| DATE | GALILEO | SCHEINER |
|---|---|---|
| 16/3/1612 | 1 | 5 |
| 17/3/1612 | 4 | 5 |
| 18/3/1612 | 4 | 3 |
| 20/3/1612 | 2 | 4 |
| 21/3/1612 | 2 | 4 |
| 31/3/1612 | 3 | 3 |
| 3/4/1612 | 4 | 5 |
| 5/4/1612 | 3 | 3 |
| 6/4/1612 | 3 | 3 |
| 7/4/1612 | 3 | 3 |

Regarding the solar activity level recorded by Galileo and Scheiner, we note that the daily group number average recorded by Galileo according to V16 for the period April–August 1612 is equal to 6.39 while the daily average of the group number recorded by Galileo taking into account the observations included in V16 in addition to the observations analyzed in this work is 5.25 (February–August 1612). In the case of Scheiner, the average of the group number for the period October–December 1611 according to V16 is 3.67 and it is very similar to the group average (3.74) adding to the calculation the sunspot observations made by Scheiner analyzed in this work. We found ten days with drawings by both observers (Table 1). Generally, Scheiner recorded more groups than Galileo. The group average obtained from Scheiner's records for these ten observation days is 3.8 while that one obtained from Galileo's observations is 2.9. Scheiner observed more groups than Galileo in five days. The greatest difference in the group counting occurred on 16 March 1612 when Scheiner recorded five groups while Galileo only one. Galileo recorded more groups than Scheiner only on 18 March 1612 when he recorded four groups and Scheiner three groups. They observed the same number of groups in four observation days (31 March and 5–7 April 1612). Figure 5 shows comparisons between the sunspot drawings made by Galileo and Scheiner in those ten common observation days. We can also see that: (i) Galileo did not record several groups recorded by Scheiner on the solar limb (group "b" on 20 – 21 March and group "e" on 31 March) and the significant group "f" on 20 – 21 March, (ii) Galileo sometimes recorded one group in places where Scheiner recorded more than one (2 groups in groups "b", "c", and "d" on 21 March and one group in groups "g" and "h" recorded by Scheiner on 3 April), and (iii) Galileo recorded a few groups not recorded by Scheiner (one group on 18 March and other below group "g" recorded by Scheiner on 31 March corresponding to group "h" assigned by Scheiner the previous days).

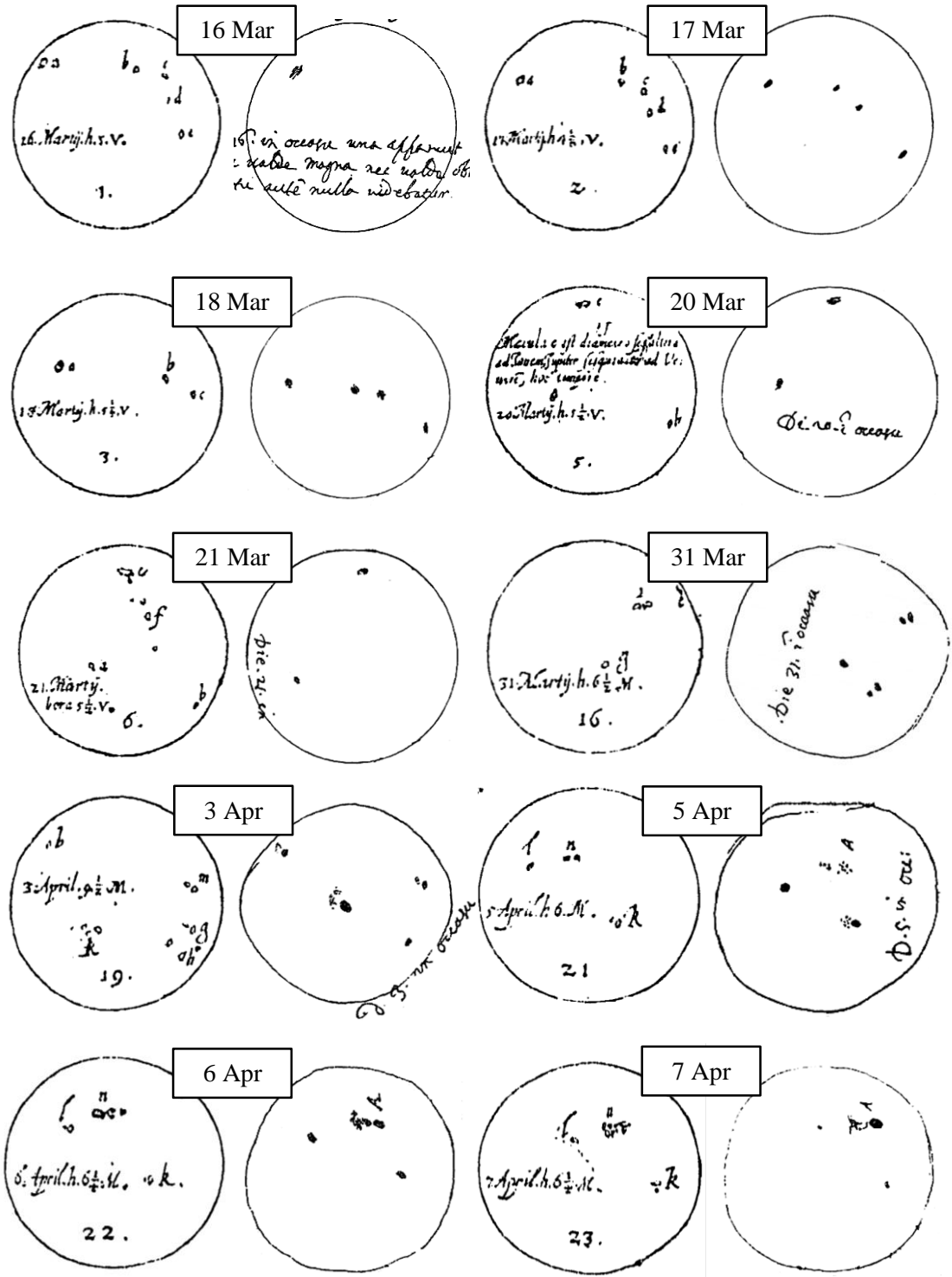

Figure 5. Comparison between the sunspot drawings made by Scheiner (drawings at first and third columns) and Galileo (drawings at second and fourth columns) for the ten common observation days. Note that some Galileo's drawings are rotated in order to show the same orientation as Scheiner's drawings [Source: Galilei 1895a].

Although Scheiner generally recorded more groups than Galileo, we want to highlight a comment made by Scheiner assuming that Galileo's observations were better. The possible reasons of this fact, according to Scheiner, could be several factors such as the telescope used by Galileo or even the sight of Galileo. Scheiner compared his observations of the greatest sunspot group observed for the period 5 – 7 April 1612 with those made by Galileo and concluded that his records were similar to those by Galileo as well as the proportion of sizes and shapes of sunspots but he found differences in the accuracy for single sunspots (Galilei & Scheiner 2010, p. 225): "… From this it is clear that Galileo does not disagree with me at all about the main shapes and the conformation of all the spots with respect to each other, but only departs from me somewhat in the precision suitable for single spots. This could result either from the strength of the light or from the shortcomings of the tube, or from the intervention of the medium, or, finally, from weakness of the eyes…". Figure 6 shows a comparison between the records made by Galileo (group A, B, and C) and Scheiner (red squares in the drawings) of the greatest sunspot group observed for the period 5 – 7 April 1612. We can see that the detail level provided by Galileo for that group is greater than that found in the sunspot drawings made by Scheiner.

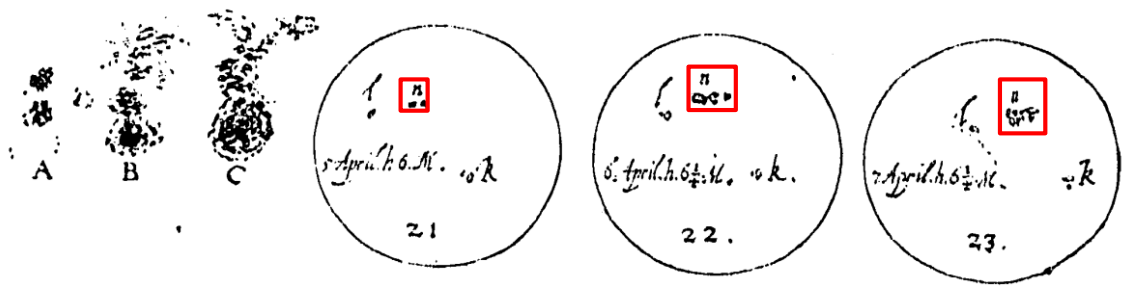

Figure 6. Comparison of the greatest sunspot group observed for the period 5 – 7 April 1612 according to Galileo (group A, B, and C) and Scheiner (groups in the red squares in the sunspot drawings) [Source: Scheiner 1612].

We can also compare some of the sunspot observations made by Scheiner and Galileo not included in V16 analyzed in this work with records made by other astronomers who observed in the same observation days as Galileo and Scheiner (Table 2).

Table 2. Number of sunspot groups recorded by Galileo (G), Scheiner (S) according to this work, and Harriot (H), Cigoli (CI), Jungius (J), and Colonna (CO) included in V16 for common observations days. The maximum daily number of groups for each day is indicated in bold when there are different number of groups recorded each day.

| DATE | G | S | H | CI | J | CO | DATE | G | S | H | CI | J | CO |
|---|---|---|---|---|---|---|---|---|---|---|---|---|---|
| 10/12/1611 |  | 5 | 3 |  |  |  | 21/3/1612 | 2 | **4** | 3 | **4** |  |  |
| 13/12/1611 |  | **4** | 3 |  |  |  | 22/3/1612 |  | 1 | 2 | **4** |  |  |
| 14/12/1611 |  | **4** | 2 |  |  |  | 23/3/1612 |  | **2** | 1 | **2** |  |  |
| 17/12/1611 |  | **5** | 2 |  |  |  | 24/3/1612 |  | **2** |  | 1 |  |  |
| 18/12/1611 |  | **5** | 2 |  |  |  | 26/3/1612 |  | 2 | 2 |  |  |  |
| 24/12/1611 |  | **3** | 1 |  |  |  | 27/3/1612 |  | 2 | 2 |  |  |  |
| 25/12/1611 |  | 2 | **3** |  |  |  | 28/3/1612 |  | **3** | 2 |  |  |  |
| 31/12/1611 |  | **5** | 3 |  |  |  | 31/3/1612 | 3 | 3 | 3 |  |  |  |
| 4/1/1612 |  | **4** | 2 |  |  |  | 1/4/1612 |  | **3** | 2 |  |  |  |
| 5/1/1612 |  | **4** | 3 |  |  |  | 3/4/1612 | 4 | **5** | 4 |  |  |  |
| 11/1/1612 |  | 3 | 3 |  |  |  | 4/4/1612 |  | 3 | 3 |  |  |  |
| 12/2/1612 | 3 |  | 3 |  |  |  | 6/4/1612 | 3 | 3 | 3 |  |  |  |
| 17/2/1612 | 1 |  | 1 |  |  |  | 7/4/1612 | 3 | 3 | 3 |  |  |  |
| 23/2/1612 | 6 |  |  | 6 |  |  | 10/4/1612 | 2 |  | 2 |  |  |  |
| 2/3/1612 | 0 |  | 0 | 0 |  |  | 16/4/1612 | **1** |  | 0 |  |  |  |
| 4/3/1612 | 0 |  | 0 | 0 |  |  | 29/4/1612 | 1 |  | **2** | 1 |  |  |
| 16/3/1612 | 1 | **5** |  | 4 |  |  | 30/4/1612 | 1 |  | **2** | 1 |  |  |
| 17/3/1612 | 4 | **5** |  | 4 |  |  | 1/5/1612 | **2** |  | **2** | 1 |  |  |
| 18/3/1612 | **4** | 3 |  | 2 |  |  | 3/5/1612 | 3 |  | **4** | 3 |  |  |
| 19/3/1612 |  | **4** | 1 | 2 |  |  | 27/11/1612 |  | 5 | **6** |  |  |  |
| 20/3/1612 | 2 | **4** | 2 | 2 |  |  | 1/8/1613 |  | **6** |  |  | 2 | 3 |

Regarding these observations, Galileo recorded sunspot drawings in 16 observation days when Harriot also observed and 12 in the case of Cigoli. Taking into account only the observation days when Galileo and Harriot observed the same day: i) the average of the sunspot group number recorded by Harriot is 2.1 while it is 1.9 according to Galileo, and ii) Harriot observed a greater number of groups than Galileo in four days and, on the contrary, Galileo in only one In addition, Harriot recorded two groups on 1 March 1612 when Galileo reported in a note that he observed sunspots but without specifying the number of groups or single sunspots. In the case of the comparison between Galileo and Cigoli regarding only days when both astronomers observed: i) the group average calculated from Cigoli's records is 2.3 while it is 2.2 from Galileo's observations, and ii) Galileo observed more groups than Cigoli in two days and Cigoli recorded more groups in other two days. We highlight one comment recorded by Cigoli in a letter sent to Galileo in March 1612 (Galilei 1895b, p. 287) to show him his sunspot observations: (English translation) "I do not think I wrote to Your Excellence that I have a telescope, and it's very good, so much that I see the clock of Saint Peter's and the hand of the clock from Santa Maria Maggiore, but I do not see the numbers as clearly as with your telescope; but if you can give me some advice for more accurate, let me know". Although Cigoli recorded observations similar to Galileo regarding the number of groups and even more groups in two observation days, the telescope used by Galileo to observe sunspots was better than that employed by Cigoli according to Cigoli's comment. From the change in the observation method by Galileo on 3 May 1612, we can see significant differences in the number of groups recorded by both observers. Thus, on 3 and 6 May 1612, Cigoli recorded 3 and 4 groups while Galileo observed 9 and 7 groups, according to V16, using the new methodology. Furthermore, in the common eight observation days when Galileo, Harriot, and Cigoli recorded observations, we can find differences in the group number recorded by the three observers in five days. Moreover, the daily average of the group number recorded by Harriot in those eight days (1.9) was higher than that by Galileo (1.4) and Cigoli (1.5).

Scheiner also made observations in observation days when Harriot and Cigoli observed. Scheiner and Harriot have 26 observation days in the same date while Cigoli and Scheiner 9 days. Regarding the observation days when Scheiner and Harriot observed the same day: i) the average of the group number recorded by Harriot in the matching 26 observation days is 2.5, significantly lower than the average of the group number according to Scheiner's records that it is 3.5, and ii) Scheiner recorded in 16 of the 26 observation days in common a greater number of groups than Harriot while Harriot only in three days recorded one group more than Scheiner In the case of the comparison with Cigoli: i) the average of the group number in accordance with Cigoli's records regarding those 9 days is 2.8 and 3.3 according to Scheiner, and ii) Cigoli recorded more groups than Scheiner on 22 March 1612, and Scheiner recorded more groups than Cigoli in five different days. The observation day number in which Harriot, Cigoli and Scheiner observed the same day was five and the daily average of the group number recorded by each one was 1.8, 2.8, and 3.0, respectively. Furthermore, according to V16, Jungius observed five groups on 6 January 1613 when Scheiner recorded 6 groups (groups A, B, C, D-E-F, G, and H in Figure 2). We note that one observation made by Scheiner on 5 January 1613 recording five groups is included in V16. After studying the sunspot observations made by Scheiner, we think that this last sunspot record could be the same record presented in this work but with a misreading by V16 in the observation day. In addition, Scheiner, Colonna, and Jungius observed on 1 August 1613. Figure 7 (top panel) shows the sunspot drawings made by Scheiner and Colonna for that date. Although we can count six groups from Scheiner's drawing and, according to V16, Colonna recorded 3 groups, actually Colonna only recorded two groups less than Scheiner: one group next to facula "a" on the solar limb and other with two sunspots closer to the apparent solar center. We also note that Jungius recorded two groups that day according to V16.

We can compare the observations made by Galileo, Scheiner, Cigoli, and Harriot only in one common observation day (21 March 1612). We have available the sunspot drawings made by Galileo, Scheiner, and Cigoli, but not that one by Harriot. Figure 7 (bottom panel) shows the sunspot drawings made by Cigoli, Galileo, and Scheiner on 21 March 1612. The group number recorded by Cigoli and Scheiner is the same (four groups) while Galileo only recorded two groups. We can see the sunspot positions recorded by Cigoli are slightly displaces to those made by Scheiner and Galileo, in particular, the sunspot corresponding to group "a" assigned by Scheiner. We note that Harriot observed three sunspot groups in this day according to V16.

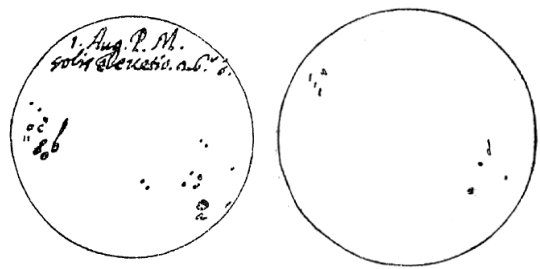

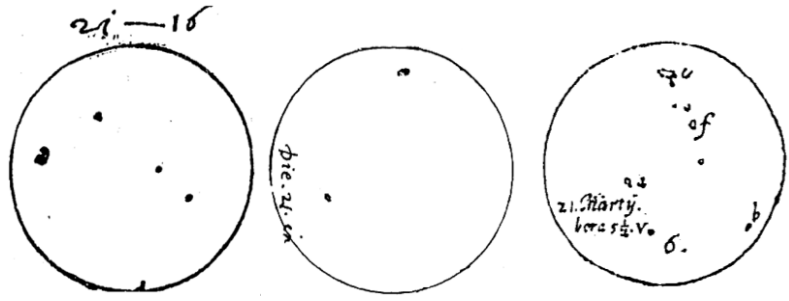

Figure 7. (Top panel) Sunspot drawings made by Scheiner (left) and Colonna (right) on 1 August 1613. Note that: (i) marks "a", "b", and "c" in Scheiner's drawing are facular regions and (ii) Colonna's drawing was rotated to show sunspots in the same orientation as Scheiner [Source: Galilei 1895b]. (Bottom panel) Sunspot drawings made by Cigoli (left), Galileo (middle), and Scheiner (right) on 21 March 1612. Note that Galileo's drawing was rotated to show sunspots with the same orientation as the other drawings [Source: Galilei 1895a,b].

## 4. Conclusions

Galileo Galilei and Christoph Scheiner carried out sunspot observations during the first years of the telescopic era generating an intense debate about the nature of this phenomenon. Some of their sunspot records were not included in the sunspot group databases (Hoyt & Schatten 1998; Vaquero et al. 2016). We here analyze these sunspot observations in order to incorporate them into V16. We can fill 18 observation days in V16 where no record is available. Moreover, we can extend the observational coverage: (i) for Scheiner, from 33 records for the period October – December 1611 included in V16 to 78 until August 1613 and (ii) for Galileo, from 51 records included in V16 for the period May – August 1612 to 72 records regarding his sunspot records from February 1612. Furthermore, the values corresponding to the group number recorded for Scheiner on 13, 29 and 30 December 1611 and Galileo on 26 April 1612 must be corrected according to this new analysis.

We have detected changes in the quality of the sunspot observations made by Galileo and Scheiner. We found a significant improvement in the quality of Galileo's sunspot drawings from 3 May 1612. This fact was because he changed the observation method from observing sunspots through the telescope to drawings where sunspots were recorded by projecting the Sun on a sheet. We can see that this fact had an impact in the number of sunspot groups recorded by Galileo by comparing his sunspot observations with Harriot's and Cigoli's records. According to the comparison to Harriot: (i) the daily average of the number of groups recorded by Galileo and Harriot from February to 3 May in the 16 common observation days is 1.9 and 2.1, respectively, while it is 6.8 and 4.2 from 3 May to August (24 common observation days) according to V16, respectively. In the case of Scheiner, we can also see a remarkable change in the sunspot drawings made in 1620s published by Scheiner (1630) in *Rosa Ursina* and his first sunspot records from 1611 to 1613 analyzed in this work. This was also due to a change in Scheiner's methodology. Scheiner passed to observe the Sun through the telescope with colored glasses in his first drawings to adopt the projection method in 1620s. Note that: (i) Carrasco et al. (2019c) indicated that Scheiner used the helioscope in 1620s, which allowed him to change his methodology and to improve the sunspot observations, and (ii) Arlt et al. (2016) pointed out the telescope used by Scheiner in 1620s had a Keplerian setup providing better image quality. Thus, this significant improvement in the quality of the observations of both astronomers affected to their group counts.

Scheiner generally recorded more groups than Galileo, Harriot, and Cigoli who recorded a similar number of groups. We concluded that the differences found in these comparisons were mainly because of the observational method employed by the observers. While Scheiner observed the Sun through colored glasses, the others astronomers observed when the Sun was, for example, behind clouds or through misty air (Galilei 1895a,b; Vokhmyanin, Arlt & Zolotova 2020). On 3 May 1612, Galileo started to use a new methodology to observe the Sun by projecting the solar image on a sheet and it improved his capacity to record sunspots. This fact has been proven by comparing the number of sunspot groups recorded by Galileo with those by Harriot and Cigoli before and after on 3 May 1612, date from which Galileo significantly recorded more groups than Cigoli and Harriot. Thus, the comparison carried out in this work can help to redefine the scale of the group number series over this important part of the 400-year sunspot record. The sunspot observations made by Galileo observing directly by telescope (before 3 May 1612) and those made by projection (from 3 May 1612) should be considered like two independent observational sets. Thus, we will avoid problems of inhomogeneity. Moreover, the sunspot observations made by Galileo on 3 May 1612 from both methodologies could be used to calibrate both observational sets. Similarly, the sunspot observations made by Scheiner in 1610s and those made in 1620s should be considered as independent for the future reconstruction of the group number index.

This work shows up the need to continue the review work of the available information about the past solar activity. We have shown that the observations here analyzed can increase the observational coverage in the earliest period of the telescopic era but this coverage must be improved because the number of records in that period is still scarce. Other documentary sources with new information should be studied. Thereby, we will be able to improve our knowledge about long-term solar activity and to make more reliable future predictions.

**Acknowledgements**

This research was supported by the Economy and Infrastructure Counselling of the Junta of Extremadura through project IB16127 and grant GR18097 (co-financed by the European Regional Development Fund) and by the Ministerio de Economía y Competitividad of the Spanish Government (CGL2017-87917-P). Authors acknowledge to the Lincean Academy Archive the information provided about the sunspot observations made by Scheiner in March–April 1612.

**Disclosure of Potential Conflicts of Interest** The authors declare that they have no conflicts of interest.

**References**

Arlt, R., Vaquero, J.M., 2020, Historical sunspot records, Living Rev. Solar Phys., 17, 1. DOI: 10.1007/s41116-020-0023-y.

Arlt R., Senthamizh Pavai V., Schmiel C., Spada F., 2016, Sunspot positions, areas, and group tilt angles for 1611-1631 from observations by Christoph Scheiner, A&A, 595, A104. DOI: 10.1051/0004-6361/201629000.

Carrasco V.M.S., 2019, Improving sunspot records: misreading of 'Rosa Ursina' by Scheiner, Observatory, 139, 153.

Carrasco V.M.S., Vaquero J.M., Gallego M.C., Villalba Álvarez J., Hayakawa H., 2019a, Two debatable cases for the reconstruction of the solar activity around the Maunder Minimum: Malapert and Derham, MNRAS: Letters, 485, L53. DOI: 10.1093/mnrasl/slz027.

Carrasco V.M.S., Gallego M.C., Villalba Álvarez J., Vaquero J.M., 2019b, Sunspot observations by Charles Malapert during the period 1618–1626: a key data set to understand solar activity before the Maunder minimum, MNRAS, 488, 3884. DOI: 10.1093/mnras/stz1867.

Carrasco V.M.S., Vaquero J.M., Gallego M.C., Muñoz-Jaramillo, A., de Toma G., Galaviz P., Arlt R., Senthamizh Pavai V., Sánchez-Bajo F., Villalba Álvarez J., Gómez J.M., 2019c, Sunspot Characteristics at the Onset of the Maunder Minimum Based on the Observations of Hevelius, ApJ, 886, 18. DOI: 10.3847/1538-4357/ab4ade.

Casas R., Vaquero J.M., Vázquez M., 2006, Solar Rotation in the 17th century, Solar Phys., 234, 379. DOI: 10.1007/s11207-006-0036-2.

Clette F., Lefèvre L., 2016, The New Sunspot Number: Assembling All Corrections, Solar Phys., 291, 2629. DOI: 10.1007/s11207-016-1014-y.

Clette F., Svalgaard L., Vaquero J.M., Cliver E.W., 2014, Revisiting the Sunspot Number. A 400-year perspective on the solar cycle, Space Sci. Rev., 186, 35. DOI: 10.1007/s11214-014-0074-2.

Eddy, J.A., 1976, The Maunder Minimum, Science, 192, 1189. DOI: 10.1126/science.192.4245.1189.

Engvold O., Zirker J.B., 2016, The Parallel Worlds of Christoph Scheiner and Galileo Galilei, J. Hist. Astron., 47, 332. DOI: 10.1177/0021828616662406.

Galilei G., 1613, Istoria e dimostrazioni intorno alle macchie Solari, Mascardi, Rome.

Galilei G., 1895a, Le opere de Galileo Galilei vol. 5, G. Barbèra, Florence.

Galilei G., 1895b, Le opere de Galileo Galilei vol. 11, G. Barbèra, Florence.

Galilei G., Scheiner C., 2010, On sunspots, Galileo Galilei and Christoph Scheiner, University of Chicago Press, Chicago.

Hoyt D.V., Schatten K.H., 1998, Group Sunspot Numbers: A New Solar Activity Reconstruction, Solar Phys., 179, 189. DOI: 10.1023/A:1005007527816.

Kopp G., Krivova N., Wu C.J., Lean J., 2016, The Impact of the Revised Sunspot Record on Solar Irradiance Reconstructions, Solar Phys., 291, 2951. DOI: 10.1007/s11207-016-0853-x.

Malapert C., 1633, Austriaca sidera heliocyclia astronomicis hypothesibus illigata, Baltazaris Belleri, Douai.

McIntosh P.S., 1990, The classification of sunspot groups, Solar Phys., 125, 251. DOI: 10.1007/BF00158405.

Muñoz-Jaramillo A, Vaquero J.M., 2019, Visualization of the challenges and limitations of the long-term sunspot number record, Nature Astron., 3, 205, DOI: 10.1038/s41550-018-0638-2.

Neuhäuser R., Neuhäuser D.L., 2016, Sunspot numbers based on historic records in the 1610s: Early telescopic observations by Simon Marius and others, Astron. Nachr., 337, 581. DOI: 10.1002/asna.201512292.

Sakurai K., 1980, The Solar Activity in the Time of Galileo, J. Hist. Astron., 11, 164. DOI: 10.1177/002182868001100302.

Scheiner C., 1612, De maculis solaribus et stellis circa Jovem errantibus accuratior disquisitio, Augustae Vindelicorum, Ausburg.

Scheiner C., 1615, Sol ellipticus: hoc est novuum et perpetuum Solis contrahi soliti Phaenomenon, quod noviter inventum Strenae loco, C. Mangius, Augsburg.

Scheiner C., 1630, Rosa Ursina Sive Sol, Andrea Fei, Bracciano.

Usoskin I.G., 2017, A history of solar activity over millennia, Living Rev. Solar Phys., 14, 3. DOI: 10.1007/s41116-017-0006-9.

Usoskin, I.G., Arlt, R., Asvestari, E., et al., 2015, The Maunder minimum (1645–1715) was indeed a Grand minimum: A reassessment of multiple datasets, A&A, 581, A95. DOI: 10.1051/0004-6361/201526652.

Vaquero J.M., Vázquez M., 2009, The Sun Recorded Through History, Springer, Berlin. DOI: 10.1007/978-0-387-92789-3.

Vaquero J.M., Gallego M.C., García J.A., 2002, A 250-year cycle in naked-eye observations of sunspots, Geophys. Res. Lett., 29, 58. DOI: 10.1029/2002GL014782.

Vaquero J.M., Svalgaard L., Carrasco V.M.S., Clette F., Lefèvre L., Gallego M.C., Arlt R., Aparicio A.J.P., Richard J.-G., Howe R., 2016, A Revised Collection of Sunspot Group Numbers, Solar Phys., 291, 3061. DOI: 10.1007/s11207-016-0982-2.

Vokhmyanin M.V., Zolotova N.V., 2018, Sunspot Positions and Areas from Observations by Galileo Galilei, Solar Phys., 293, 31. DOI: 10.1007/s11207-018-1245-1.

Vokhmyanin M.V., Arlt, R., Zolotova N.V., 2020, Sunspot Positions and Areas from Observations by Thomas Harriot, Solar Phys., 295, 39. DOI: 10.1007/s11207-020-01604-4.

Yau K.K.C., Stephenson F.R.: 1988, A revised catalogue of Far Eastern observations of sunspots (165 BC to AD 1918), Quart. J. Roy. Astron. Soc., 29, 175.